\documentclass[12pt]{elsart}
\usepackage{graphicx}
\usepackage{amssymb}
\usepackage{amsmath}

\begin{document}
\renewcommand{\thefootnote}{\fnsymbol{footnote}} 
\renewcommand{\theequation}{\arabic{section}.\arabic{equation}}

\journal{Phys. Lett. A}

\begin{frontmatter}
\title{Modeling vacancies and hydrogen impurities in graphene: A molecular
point of view}

\author[CTChem]{G. Forte}
\author[CTChem]{A. Grassi}
\author[CTChem]{G. M. Lombardo}
\author[IMM]{A. La Magna}
\author[CTPhys,INFN,CNISM]{G. G. N. Angilella\thanksref{corr}}
\author[CTPhys,CNISM]{R. Pucci}
\author[CTPhys]{R. Vilardi}

\address[CTChem]{Dipartimento di Scienze Chimiche, Facolt\`a di Farmacia,
   Universit\`a di Catania,\\
Viale A. Doria, 6, I-95126 Catania, Italy}
\address[IMM]{IMM, CNR, Catania, Italy}

\address[CTPhys]{Dipartimento di Fisica e Astronomia, Universit\`a di
   Catania,\\ Via S. Sofia, 64, I-95123 Catania, Italy}
\address[INFN]{INFN, Sez. Catania, Italy}
\address[CNISM]{CNISM, UdR di Catania, Italy}
\thanks[corr]{Corresponding author. E-mail: {\tt giuseppe.angilella@ct.infn.it}.}
\date{\today}

\begin{abstract}
We have followed a `molecular' approach to study impurity effects in graphene. 
This is thought as the limiting case of an infinitely large cluster of benzene
rings. Therefore, we study several carbon clusters, with increasing size, from
phenalene, including three benzene rings, up to coronene~61, with 61 benzene
rings. The impurities considered were a chemisorbed H atom, a vacancy, and a
substitutional proton. We performed HF and UHF calculations using the STO-3G
basis set. With increasing cluster size in the absence of impurities, we find a
decreasing energy gap, here defined as the HOMO-LUMO difference. In the case of
H chemisorption or a vacancy, the gap does not decrease appreciably, whereas it
is substantially reduced in the case of a substitutional proton. The presence of
an impurity invariably induces an increase of the density of states near the
HOMO level. We find a zero mode only in the case of a substitutional proton. In
agreement with experiments, we find that both the chemisorbed H, the
substitutional proton, and the C atom near a vacancy acquire a magnetic moment.
The relevance of graphene clusters for the design of novel electronic devices is
also discussed.

\end{abstract} 

\end{frontmatter}

\section{Introduction}

Since the pioneering works of Wallace \cite{Wallace:47}, Coulson
\cite{Coulson:52a}, and Semenoff \cite{Semenoff:54}, several authors have
studied the peculiar electronic and structural properties of graphene, a single
sheet of graphite, originally only considered as a prototype of the graphite's
surface. A breakthrough in these studies arrived with the experimental
realization of graphene \cite{Novoselov:04}. The increasing interest in this
material stems both from the analogy of its behavior with many phenomena of
quantum electrodynamics (QED) \cite{Katsnelson:07a}, and because of its
prospective applications in the fabrication of novel electronic devices. To this
aim, it is important to have systems with a tunable band gap. We will stress
that the graphene clusters (GC) here considered, even if they are characterized
by a zig-zag boundary, have a semiconducting behavior, at variance with zig-zag
graphene nanoribbons (ZGNR), which have a metallic behavior \cite{Yan:07}. The
gap strongly depends on the cluster size, so that it will be highly desirable to
have method allowing to control the size of graphene quantum dots.

Graphene is a robust material. However, defects, such as ripples, adatoms,
vacancies and charges induced by the substrate, can modify its electronic
properties \cite{CastroNeto:08}. In the past, the study of vacancies and
impurities in carbon-based systems has been important for several aspects.

(i) Graphene nanostructures are believed to be attractive structures for
hydrogen storage. Preliminary results in this direction have been reported in
the case of single-walled carbon nanotubes (SWCNT) \cite{Dillon:97}.

(ii) It has been found \cite{Esquinazi:03} that proton irradiation of highly
oriented pyrolitic graphite samples triggers ferro- or ferrimagnetism. It is now
commonly accepted that not only vacancies, but also hydrogen chemisorbed on
graphene can acquire a magnetic moment.

(iii) From an astrophysical point of view, it would be of interest to understand
why H$_2$ is the most abundant molecule in the interstellar medium (ISM),
despite the continuous dissociation of molecular hydrogen by UV radiation and
cosmic rays \cite{Hornekaer:06}.

Of course, there can be relevant differences in the characterization of defects
in graphite, nanotubes, or graphene. For instance, if one considers the
screening of a charged impurity, Cheianov and Fal'ko \cite{Cheianov:06} have
shown that the electron density $\delta\rho$ in graphene decays as
$\delta\rho(r) \sim r^{-3}$ at long distances $r$ from the impurity. Such a
behavior should be contrasted with that derived by Lau and Kohn \cite{Lau:78}
for a two-dimensional (2D) electron gas, where $\delta\rho(r) \sim r^{-2}$. The
reason of such a difference is due to the peculiar chiral properties of the
electrons in graphene. It has been suggested \cite{Bena:07} that such a
difference can be used to distinguish between monolayer and bilayer graphene in
Fourier transformed scanning tunneling spectroscopy (FTSTS). Actually, impurity
induced quantum interferences have been used in scanning tunneling microscopy
(STM) \cite{Mallet:07} to identify experimentally mono- or bilayer graphene.
Previously, STM was used \cite{Ruffieux:00,Ruffieux:05} to study chemisorption
of hydrogen on the basal plane of graphite and atomic vacancy formation.

From a theoretical point of view, the effects on the electronic structure of
graphene due to the presence of vacancies, local impurities, and substitutional
impurities have been studied by Pereira \emph{et al.} \cite{Pereira:08}, by
using a tight-binding Hamiltonian with a local potential $U$. These authors have
found general trends in the low-energy spectrum of graphene connected with
localized zero modes, resonances and gaps induced by the above mentioned
defects. However, in some cases, it is important to have a more detailed
description of the local interaction of the defects with graphene. For this
reason, in the present work, we will make recourse to molecular models,
which in the past have been used to study the chemisorption of hydrogen on
graphite surface \cite{Klose:92,Fromherz:93}. According to these calculations,
the graphite surface is modeled as a cluster of benzene rings, whose external
dangling bonds are saturated by hydrogen atoms. The most frequently used model
clusters are polycyclic aromatic hydrocarbons (PAH), and particularly coronene
(C$_{24}$H$_{12}$). More recently, this model cluster has been used by
Patchkovskii \emph{et al.} \cite{Patchkovskii:05} to study the interaction of
molecular hydrogen with graphene nanostructures. These authors have performed
calculations at the level of second-order M\o{}ller-Plesset (MP2) perturbation
theory \cite{Moeller:34}.

In the present work, we will use such model clusters to simulate graphene, at
the level of \emph{ab initio} Hartree-Fock (HF) and unrestricted \emph{ab
initio} Hartree-Fock (UHF) theory. In some cases, also MP2 calculations will be
presented, but the main emphasis will be more on the trends obtained by
increasing the cluster size, than in the study of correlation effects. Actually,
Martin \emph{et al.} \cite{Martin:08} have argued that correlation effects are
small in graphene. A similar result was found by Siringo \cite{Siringo:84} in
the study of H chemisorbed on graphene \cite{note:Siringo}. Furthermore,
recently the Manchester group has found that the electron energy levels in
quantum dots in graphene follow a statistics characteristic of `Dirac
billiards', rather than a Poisson distribution, as is expected for an infinite
system \cite{Ponomarenko:08}: therefore, quantization effects are quite
important. It is interesting, in this context, to derive these results also
within a molecular approach. In the present work, we have used clusters up to 61
benzene rings, and we have considered the cases of a single vacancy, of a
chemisorbed H atom, and of a substitutional proton.

The paper is organized as follows. In Sec.~\ref{sec:models}, we describe the
models used. In Sec.~\ref{sec:results} we present our results and a comparison
thereof with density functional theory (DFT) and other types of calculations.
Finally, in Sec.~\ref{sec:conclusions} we summarize and draw our conclusions.

\section{The model clusters}
\label{sec:models}

\begin{figure}[t]
\centering
\includegraphics[bb=253 21 609 343,clip,width=0.9\columnwidth]{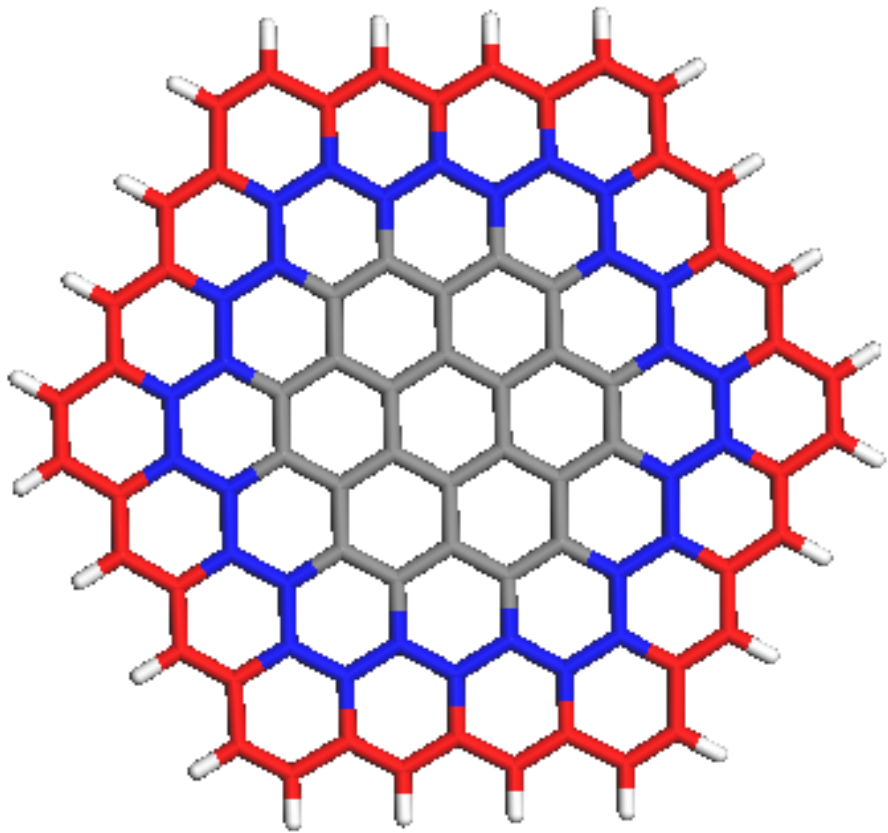}
\caption{(Color online.) Showing coronene, C$_{24}$H$_{12}$ (gray rods),
coronene~19, C$_{54}$H$_{18}$ (gray and blue rods), and coronene~37,
C$_{96}$H$_{24}$ (gray, blue, and red rods).}
\label{fig:coronene2plus}
\end{figure}

The molecular clusters which we have considered are: (i) phenalene (phenalenyl
radical); (ii) pyrene; (iii) coronene; (iv) coronene~19, \emph{viz.} coronene
surrounded by another series of benzene rings, for a total of 19 benzene rings;
(v) coronene~37, \emph{viz.} coronene surrounded by two more series of benzene
rings, for a total of 37 benzene rings; (vi) coronene~61. Coronene~37 is shown
in Fig.~\ref{fig:coronene2plus}. We believe that the larger the cluster, the
better it models graphene. For this reason, we have chosen to place in the
middle of the cluster: (a) a chemisorbed H atom; (b) a carbon vacancy; (c) a
proton in the substitutional position of a carbon atom.

For clusters (i)--(iii) we have used HF calculations with a STO-3G basis set and
HF calculations with a 6-311G basis set \cite{Frisch:04}. In these cases we have
also performed MP2 calculations. All the remaining larger clusters have been
treated at the HF level using a STO-3G basis set. In the cases of H
chemisorption, carbon vacancy and proton substitution, we have performed UHF
calculations, because in all these cases there is one $\pi$-electron missing.
The cluster geometry has been optimized in all cases.

\section{Results}
\label{sec:results}

\begin{figure}[t]
\centering
\includegraphics[width=0.9\columnwidth]{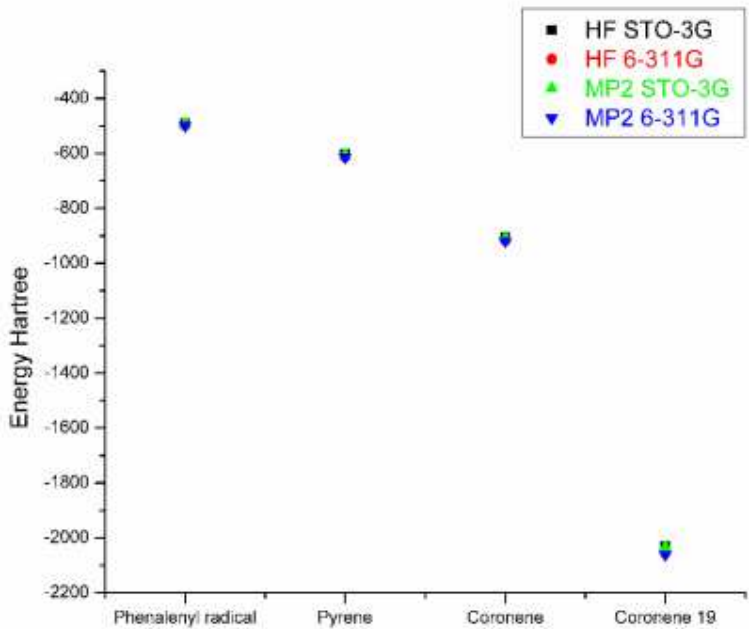}
\caption{(Color online.) Showing the dependence of the cluster energy as a
function of the cluster under consideration, for the various methods (HF or MP2)
and basis sets (STO-3G or 6-311G) employed in the present work. On this scale,
the results of different approximations are almost indistinguishable.}
\label{fig:levels}
\end{figure}

\begin{figure}[t]
\centering
\includegraphics[width=0.9\columnwidth]{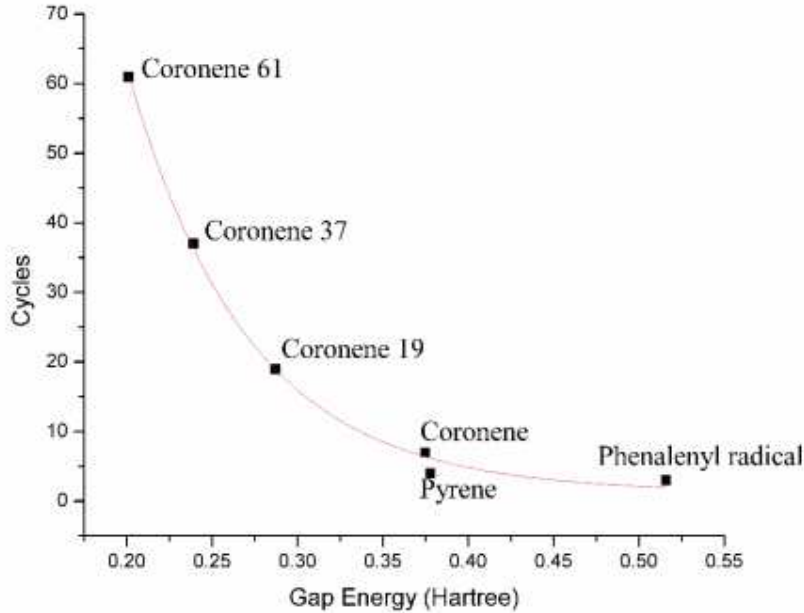}
\caption{(Color online.) Showing the dependence of the HOMO-LUMO difference (gap
energy) \emph{vs} the number of carbon cycles for the various clusters studied
in the present work. The dashed line is a guide to the eye.}
\label{fig:HOMO-LUMO}
\end{figure}

\subsection{Pure graphene clusters}
\label{sec:puregraphene}

First of all, we note that the total energy as a function of the number of
benzene rings in the clusters (cf. Fig.~\ref{fig:levels}) depends weakly
(\emph{i.e.} to within 2~\%) on the basis set or on the specific approximation
employed (HF, MP2). This lends support to our conviction that relevant
conclusions can be drawn already for one of the largest cluster considered,
\emph{i.e.} coronene~37, within the framework of the HF approximation with the
STO-3G basis set.

Secondly, we note that the gap, which in these kind of calculations is described
by the HOMO-LUMO difference, is decreasing by increasing the size of the
cluster, as shown in Fig.~\ref{fig:HOMO-LUMO}. This is in agreement with the
consideration that our clusters resemble more closely graphene, as the number of
benzene rings increases. The smallest gap which we have obtained is 5.48~eV.

Extrapolation of our data seems to indicate that the gap would close at a
cluster size of $\gtrsim 1000$~benzene rings. This is in qualitative agreement
with the experimental finding that graphene quantum dots as large as $\sim
30$~nm are characterized by a gap of $\sim 0.5$~eV \cite{Ponomarenko:08}. We
find a C--C nearest neighbours equilibrium distance $d_{\mathrm{C-C}} =
1.42$~\AA, which is stable with respect to the different approximations and
cluster sizes considered here. This in striking agreement with the typical bond
length for $sp^2$ C--C bonds for graphene and graphite, which is 1.42~\AA.

Our minimization procedure yields for the clusters considered in this case a
planar configuration, at variance with the experimental finding for graphene,
showing the presence of ripples \cite{Stolyarova:07,Meyer:07}. This discrepancy
may be related to the specific boundary conditions used in our calculations.
Such a result is not in contradiction with the known fact that a nonzero amount
of curvature is required in order to stabilize a two-dimensional crystal, as a
consequence of Mermin-Wagner theorem \cite{Peierls:34,Mermin:66,Mermin:68}.
Indeed, the systems we are considering here are not truly infinite lattices, but
rather finite-size clusters. Furthermore, we note that Choi \emph{et al.}
\cite{Choi:08}, by using a cluster with a monovacancy and 127~C atoms, and
performing spin-polarized density functional theory calculations, have found an
almost planar configuration.

\begin{figure}[t]
\centering
\includegraphics[width=0.9\columnwidth]{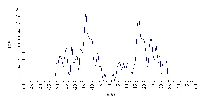}\\
\includegraphics[width=0.9\columnwidth]{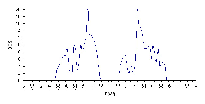}
\caption{(Color online.) Showing the DOS for pure coronene~37 (upper panel), and
for coronene~37 with a chemisorbed H atom (lower panel), as a function of
energy.}
\label{fig:DOS}
\end{figure}

\begin{figure}[t]
\centering
\includegraphics[width=0.8\columnwidth]{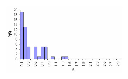}\\
\includegraphics[width=0.8\columnwidth]{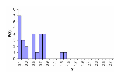}\\
\includegraphics[width=0.8\columnwidth]{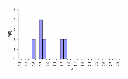}
\caption{(Color online.) Showing the level-spacing distribution $P(s)$ for
coronene~37, corresponding to different energy intervals centered around zero:
15~eV (topmost panel), 10~eV (intermediate panel), 5~eV (bottom panel). It can
be noted that $P(s)$ looses the Poissonian behavior in favor of a Gaussian
behavior as the energy interval decreases.}
\label{fig:Ps}
\end{figure}

In Fig.~\ref{fig:DOS} we report the density of states (DOS) as a function of
energy (eV) for coronene~37. We note that (i) a nearly straight line joins the
peaks near the HOMO energy: such a trend mimics the linear dependence which is
found in the DOS near the Fermi level in the infinite system
\cite{CastroNeto:08}; (ii) particle-hole symmetry is absent, as is obtained in
tight-binding calculations with a nonzero next-nearest hopping parameter
$t^\prime$ \cite{CastroNeto:08}. However, here this asymmetry is not very
pronounced than in the case $t^\prime \neq 0$. This is probably due to the fact
that in our calculations we do not restrict the interaction to next-nearest
neighbors. We have also considered the level-spacing distribution $P(s)$, where
the dimensionless spacing $s$ is defined as the energy difference $\Delta E_i =
E_i - E_{i-1}$ between successive levels, divided by the average $\langle \Delta
E_i \rangle$ of the energy difference between successive levels
\cite{DeRaedt:08}. The quantity $P(s)$ yields the number of energy differences
for which $s-\Delta/2 < \Delta E_i / \langle \Delta E_i \rangle \leq s +
\Delta/2$, where $\Delta$ is the step of the histogram. We have found that
$P(s)$ strongly depends on the energy range considered around $E=0$ (cf.
Fig.~\ref{fig:Ps}). One can also recognize a strong dependence on the cluster
size, so that no definite conclusion can be achieved at the present status of
the calculations.

\subsection{Clusters with a chemisorbed H atom}
\label{sec:chemisorbed}

\begin{figure}[t]
\centering
\includegraphics[bb=152 20 705 341,clip,width=0.9\columnwidth]{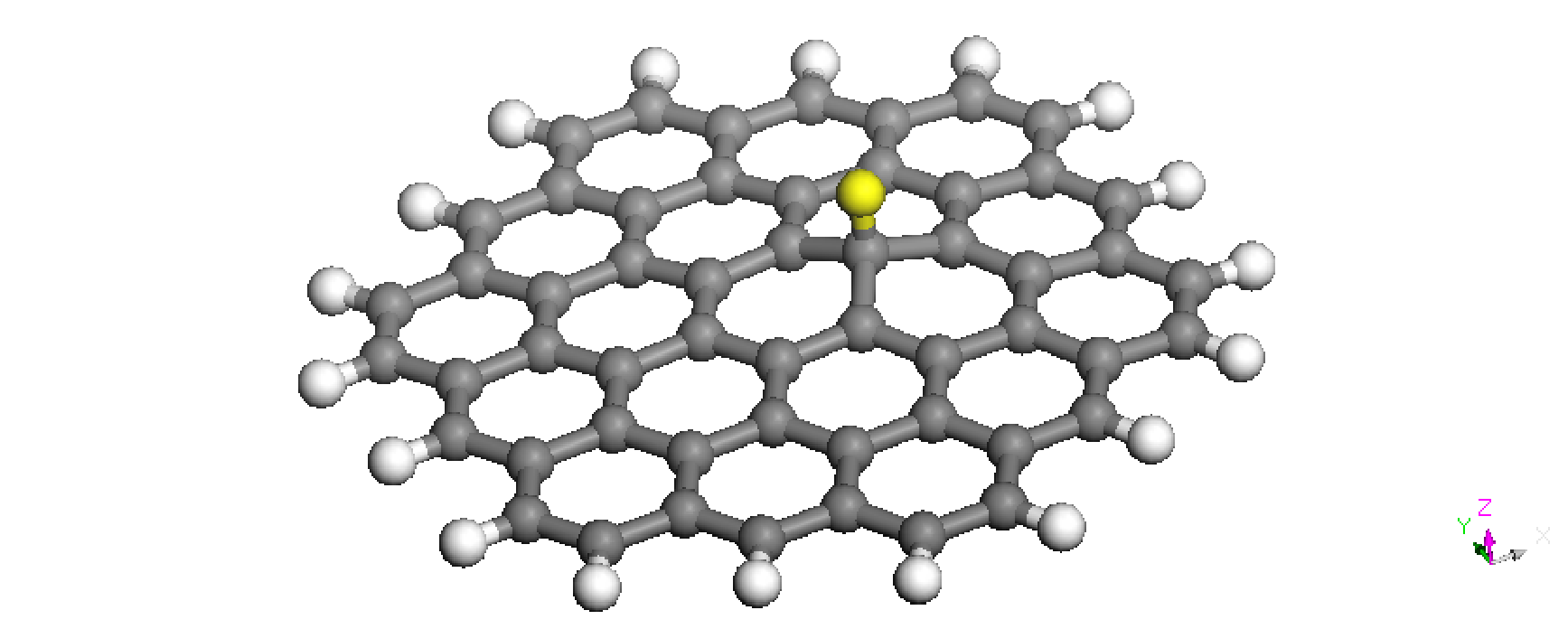}
\caption{(Color online.) Showing coronene~37 with a chemisorbed H atom.}
\label{fig:largestHatom}
\end{figure}

The largest cluster considered in the presence of a chemisorbed H atom is
reported in Fig.~\ref{fig:largestHatom}, where the H adatom is depicted with a
yellow color. 
The total energy, as a function of the cluster size
and the approximation used, is quite similar to that reported in
Fig.~\ref{fig:levels}, and will not be shown here. 

We find an average equilibrium distance of 1.53~\AA{} between two in-plane
nearest neighbor C atoms, forming an angle $\angle$HCC$=105.3^\circ$ with the H
adatom, while the C atom, on top of which the H atom resides, emerges from the
plane of 0.71~\AA. The distance of the H atom from the underneath C atom is
1.10~\AA. The distance of the C atoms surrounding the adatom is of 0.30~\AA{}
with respect to the planar configuration. These values are in qualitative
agreement with recent DFT calculations \cite{Boukhvalov:08}. We find that (i)
the most stable position is that with H on top of a carbon atom; (ii) the
underlying C atom is induced out of the plane; (iii) there are small differences
between the $\alpha$ and $\beta$ sets of eigenvalues positions. We will
therefore report only the results for the $\alpha$ set.

\begin{figure}[t]
\centering
\includegraphics[width=0.9\columnwidth]{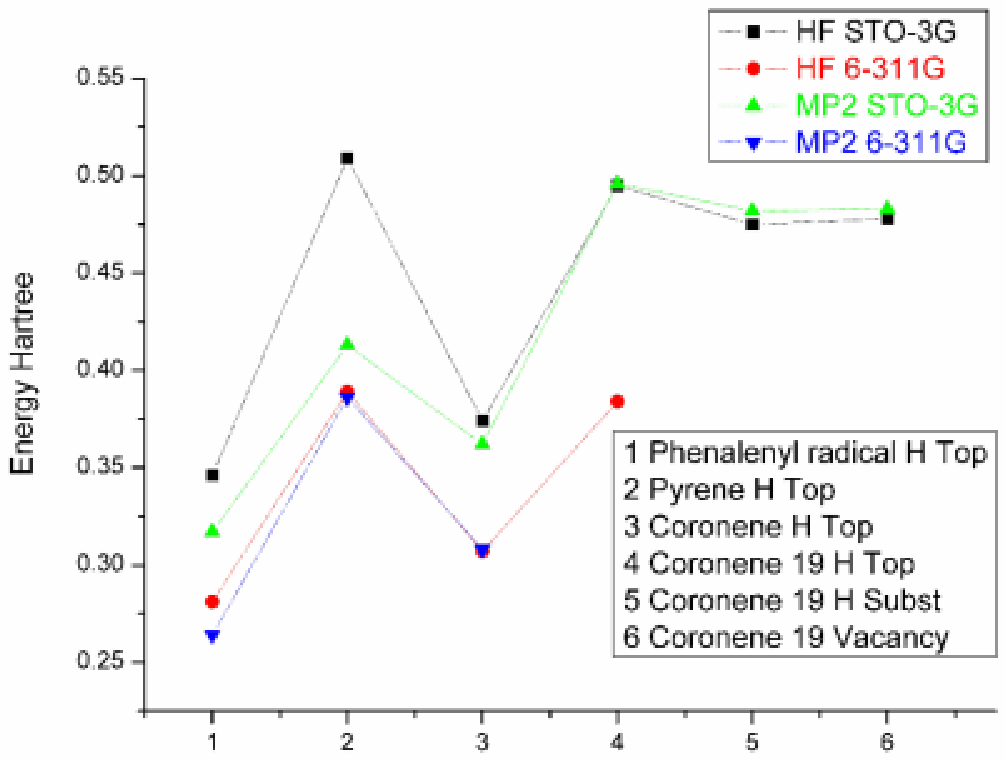}
\caption{(Color online.) Showing the dependence of the HOMO-LUMO difference (gap
energy) \emph{vs} the various clusters considered, in the presence
of a chemisorbed H atom, for the different methods and basis sets employed in
this work.}
\label{fig:rings}
\end{figure}

An important difference found in the presence of a chemisorbed H atom, with
respect to the pure graphene clusters discussed in Sec.~\ref{sec:puregraphene},
is the trend of the gap as a function of the number of the benzene rings in the
cluster, which is reported in Fig.~\ref{fig:rings}. No evidence of a gap
decrease appears in this case, which supports the results of Pereira \emph{et
al.} \cite{Pereira:08} that hydrogen chemisorption on graphene induces the
opening of a gap. Also in agreement with such findings \cite{Pereira:08} is the
behavior of the DOS, which is reported a dashed blue line in Fig.~\ref{fig:DOS}.
One can see a clear enhancement of the DOS near the HOMO level, which Pereira
\emph{et al.} \cite{Pereira:08} have found as due to a resonance induced by the
chemisorbed hydrogen. Of course, in this case we have used UHF calculations,
which allowed us to estimate also the magnetic moment acquired by the H atom,
which for a corenene~37 cluster is $-0.33$~$\mu_{\mathrm{B}}$.

The above mentioned results, in the framework here considered, do not depend
strongly on the cluster size considered, as reported in
Tab.~\ref{tab:moment}, where a comparison is presented between the results
obtained with the coronene~19 and the coronene~37 clusters, by using a STO-3G
basis set.

\begin{table}[t]
\caption{Showing values of the total energy $E$ (Hartree), HOMO energy
(Hartree), LUMO energy (Hartree), HOMO-LUMO gap (eV), and magnetic moment $\mu$
($\mu_{\mathrm{B}}$), for the coronene~19 and coronene~37 clusters, and the same
clusters with a H adatom on top, with a vacancy, with substitutional proton, and
with a substitutional H atom, using a STO-3G basis set.}
\label{tab:moment}
\begin{tabular}{lrrrrr}
\hline
 & $E$ & HOMO & LUMO & gap & $\mu$ \\
 \hline
\multicolumn{6}{l}{coronene~19} \\
pure & $-2030.796$ & $-0.159$ & $0.128$ & $7.810$ & --- \\
with H top & $-2031.701$ & $-0.252$ & $0.243$ & $13.465$ &
$-0.333$\\
with vacancy & $-1993.461$ & $-0.262$ & $0.216$ & $13.007$ & $1.838$
\\
with proton & $-1993.834$ & $-0.335$ & $-0.120$ & $5.850$ & $0.917$
\\
with H subst. & $-1994.058$ & $-0.263$ & $0.212$ & $12.925$ & $0.849$ \\
\hline
\multicolumn{6}{l}{coronene~37} \\
pure & $-3605.818$ & $-0.137$ & $0.101$ & $6.476$ & --- \\
with H top & $-3607.059$ & $-0.251$ & $0.240$ & $13.361$ &
$-0.334$\\
with vacancy & $-3568.808$ & $-0.259$ & $0.215$ & $12.898$ & $1.838$
\\
with proton & $-3569.186$ & $-0.319$ & $-0.114$ & $5.578$ & $0.833$ \\
with H subst. & $-3569.401$ & $-0.261$ & $0.212$ & $12.871$ & $0.851$ \\
\hline
\end{tabular}
\end{table}

\subsection{Clusters with a vacancy}
\label{sec:vacancy}

\begin{figure}[t]
\centering
\includegraphics[width=0.9\columnwidth]{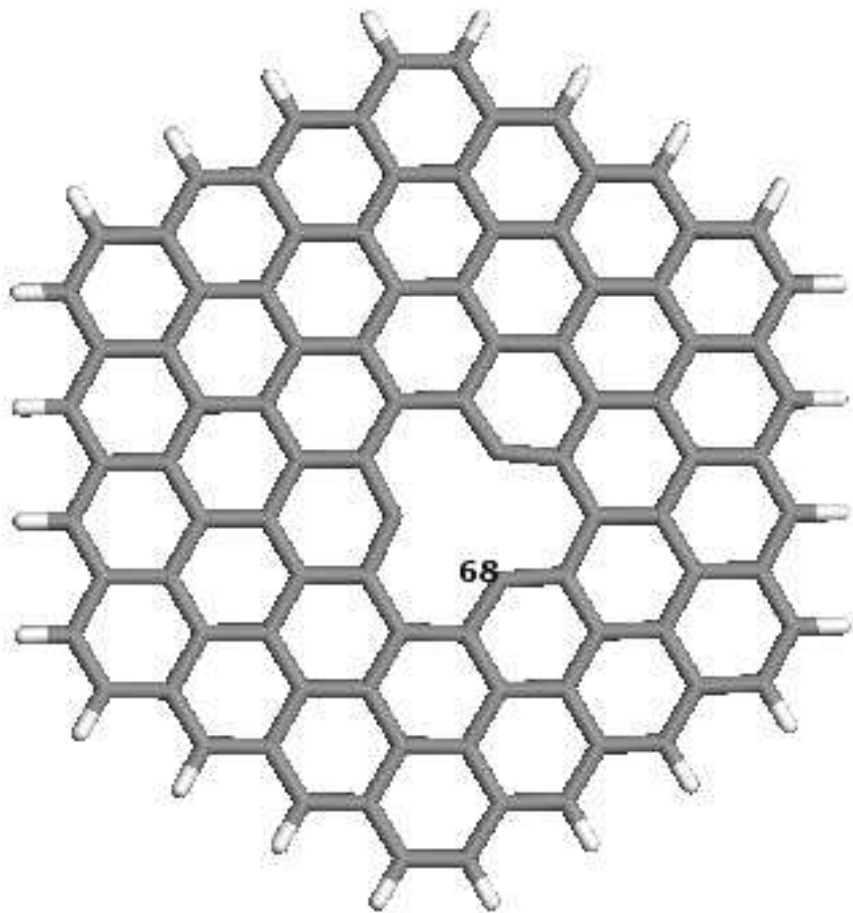}
\caption{(Color online.) Showing coronene~37 in the presence of a vacancy. Number
68 indicates the position of the C atom around the vacancy referred to in the
main text.}
\label{fig:vacancy}
\end{figure}

\begin{figure}[t]
\centering
\includegraphics[width=0.9\columnwidth]{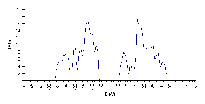}\\
\includegraphics[width=0.9\columnwidth]{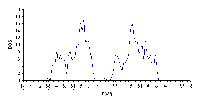}
\caption{Upper panel: Showing DOS of coronene~37 with a vacancy.
Lower panel: Showing DOS of coronene~37 with a substitutional proton.}
\label{fig:DOSvac}
\end{figure}

In Fig.~\ref{fig:vacancy} we report the clusters considered in the presence of a
vacancy. The same considerations made above for the total energy per number of
benzene rings in Sec.~\ref{sec:puregraphene} and \ref{sec:chemisorbed} above
apply to this case as well. The most stable electronic configuration is the
triplet state, and the gap has a similar trend and similar values as those
obtained for H chemisorption (cf. Tab.~\ref{tab:moment}). From
Fig.~\ref{fig:DOSvac}, one can see that also in this case the DOS presents an
increase near the HOMO level with respect to the case in the absence of the
vacancy. The main difference with respect to the case with H on top, is relative
to the magnetic moment acquired by a carbon atom near the vacancy (see atom
labelled C$^{68}$ for coronene~37 in Fig.~\ref{fig:vacancy}). The value obtained
in this case is $\approx 1.84$~$\mu_{\mathrm{B}}$, which presents just small
changes in going from coronene~19 to coronene~37, or by using the $\alpha$ or
$\beta$ sets of eigenvalues. The fact that the magnetic moment is smaller when
we consider H chemisorption instead of a vacancy is in qualitative agreement
with the results of Yaziev and Helm \cite{Yaziev:07}. These authors have found
$1$~$\mu_{\mathrm{B}}$ per hydrogen chemisorption defect, and
$1.12-1.53$~$\mu_{\mathrm{B}}$ per vacancy defect, depending on the defect
concentration. A value of $1.52$~$\mu_{\mathrm{B}}$ for a monovacancy is
reported by Choi \emph{et al.} \cite{Choi:08}. From this comparison, we can
deduce that the magnetic moments are sensitive to the model calculations
employed, but our results are in overall agreement with the most recent works.

\begin{figure}[t]
\centering
\includegraphics[width=0.9\columnwidth]{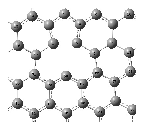}
\caption{Showing C atoms around the vacancy. The distances between the various
atoms are reported in Tab.~\ref{tab:vacdist}.}
\label{fig:vacdist}
\end{figure}

Fig.~\ref{fig:vacdist} and Tab.~\ref{tab:vacdist} report the distances between
the various C~atoms around the vacancy.

\begin{table}[t]
\caption{Distances (in \AA{}) between the various C~atoms $(i,j)$ around the vacancy, as
listed in Fig.~\ref{fig:vacdist}.}
\label{tab:vacdist}
\begin{tabular}{ccc}
\hline
$i$ & $j$ & $d_{ij}$ \\
\hline
68 & 69 & 1.38\\
68 & 70 & 1.38\\
68 & 71 & 2.39\\
68 & 66 & 2.39\\
68 & 17 & 2.77\\
68 & 6  & 2.77\\
68 & 1  & 2.63\\
68 & 5  & 2.63\\
68 & 80 & 2.41\\
68 & 82 & 2.41\\
\hline
\end{tabular}
\end{table}

\subsection{Clusters with a substitutional proton}

Last, we have considered the same clusters as in Fig.~\ref{fig:coronene2plus},
but now with a proton substituting a carbon atom at the cluster center. Such a
configuration has been realized by considering a de-electronated hydrogen atom.
The same considerations for the total energy apply also to this system, as
discussed in the presvious sections. The distances between the proton and the
nearest neighbor C atoms are 1.03~\AA{} for the C$^{68}$ atom, and 1.83~\AA{}
for the other two next nearest neighbor atoms. The cluster turns out to be flat.
The DOS presents a peak near zero energy (cf. Fig.~\ref{fig:DOSvac}), as found
by several authors when considering the effect of impurities in graphene
\cite{Pereira:08,Hu:08}. We note that, for a substitutional proton, the gap is
more than a half smaller with respect to the cases considered in
Sections~\ref{sec:chemisorbed} and \ref{sec:vacancy} (Cf.
Tab.~\ref{tab:moment}). It is even smaller with respect to a cluster with a
substitutional H, \emph{i.e.} a system presenting charge neutrality, which is
also reported in Tab.~\ref{tab:moment}. We argue that the gap is a sensitive
quantity with respect to the injections of extra charges. The magnetic moment on
the proton has an intermediate value with respect to the two cases considered
previously. It is more sensitive to the cluster size going from
$0.91$~$\mu_{\mathrm{B}}$ from coronene~19 to $0.83$~$\mu_{\mathrm{B}}$ for
coronene~37, but is similar to that obtained for the substitutional H, which is
$0.85$~$\mu_{\mathrm{B}}$ for both clusters (cf. Tab.~\ref{tab:moment}). The
present model is probably the most interesting in realtion to the experiments
performed by irradiating graphite with protons \cite{Esquinazi:03}.

\section{Conclusions}
\label{sec:conclusions}

In the present work we have followed a `molecular' approach to study impurity
effects in graphene. The impurities considered were a chemisorbed H atom, a
vacancy, and a a positively charged substitutional H. We modelled these systems
by considering clusters containing up to 37 benzene rings. We performed HF and
UHF calculations using the STO-3G basis set.

The main emphasis of this work is not on correlation effects, which in some
cases have been treated at the MP2 level, but on the size of the cluster. From
the results discussed in the present work, we believe that coronene~37 is
sufficiently large to ensure a local behavior similar to that of bulk graphene,
especially near the center of the cluster. This is supported by the small
changes in the physical properties studied here in going from coronene~19 to
coronene~37. Of course, some features differ considerably from the finite
clusters and the infinite system. For instance, we still observe quite a large
gap in the density of states of finite clusters, as compared to that of
graphene, while we have shown some important trends, which look quite promising
in interpreting the general features of graphene: (i) The gap decreases by
increasing the cluster size; (ii) The gap does not decrease if we consider H
chemisorption or a vacancy; (iii) the gap is drastically reduced if we consider
a substitutional proton. Other important features are: (i) an increase of the
DOS near the HOMO level for all the impurities considered here with respect to
the pure system; (ii) the appearance of a zero mode in the DOS if we consider a
substitutional proton.

These results are for many aspects in agreement with the theoretical work of
Pereira \emph{et al.} \cite{Pereira:08}, who have studied local disorder in
graphene by using a tight-binding approach. Such a method allowed these authors
to treat lattices with $4\times 10^6$ carbon atoms, thus reaching the size of
real samples. However, they cannot treat lattice distortions around the defect,
or the specific features connected with the particular impurity introduced.

In agreement with experiments \cite{Esquinazi:03}, we have found that
chemisorbed H, substitutional proton, and C atom near a vacancy acquire a
magnetic moment. The determination of a precise value thereof is quite delicate,
but we believe that the trend on going from one defect to another is correct, in
view of the comparison of our results with the calculations of Yazyev and Helm
\cite{Yaziev:07}.

We could not derive definite conclusions as regards the spacing of the levels
near the HOMO state, and further work is thus required in this direction. In the
future, it should be possible to treat even larger clusters, and these will be
able to treat more realistically especially graphene quantum dots. We find that
the gap strongly depends on the graphene cluster size, so that it is essential
to control the size of graphene quantum dots in order to obtain a particular gap
value. This is relevant for the design of novel electronic devices, as well as
it is important to control the impurity level.


\ack

The authors thank Professor N. H. March for useful discussions over the general
area embraced by the present paper.

\begin{small}
\bibliographystyle{mprsty}
\bibliography{a,b,c,d,e,f,g,h,i,j,k,l,m,n,o,p,q,r,s,t,u,v,w,x,y,z,zzproceedings,Angilella,notes}
\end{small}

\end{document}